\documentclass[%
reprint,
superscriptaddress,
 amsmath,amssymb,
 aps,
 pra,
]{revtex4-1}
 
\usepackage{amsmath}
\usepackage{amssymb}
\usepackage{amsfonts}
\usepackage{dsfont}
\usepackage{graphicx}
\usepackage[utf8]{inputenc}
\usepackage[T1]{fontenc}
\usepackage[english]{babel}
\usepackage{xcolor}
\usepackage{tikz}
\usepackage{tikz-feynman}

\graphicspath{{../figures/}}

\newcommand{\p}{\partial}
\newcommand{\avg}[1]{\left\langle #1 \right\rangle}

\newcommand{\ha}{\hat{a}}

\let\Re\relax
\DeclareMathOperator{\Re}{Re}

\DeclareMathOperator{\Tr}{Tr}

\begin{document}
\title{Transmission spectra of bistable systems: from ultra--quantum to classical regime}
\date{\today}
\author{Evgeny V. Anikin}
\affiliation{Skolkovo Institute of Science and Technology, 121205 Moscow, Russia}
\author{Natalya S. Maslova} 
\affiliation{
    Quantum Technology Centrum and Chair of Quantum Electronics, Department of Physics, 
    Lomonosov Moscow State University, 119991, Moscow, Russia
    }
\author{Nikolay A. Gippius}
\affiliation{Skolkovo Institute of Science and Technology, 121205 Moscow, Russia}
\author{Igor M. Sokolov}
\affiliation{Institut für Physik and IRIS Adlershof, Humboldt Universität 
             zu Berlin, Newtonstraße 15, 12489 Berlin, Germany}
\begin{abstract}
We present an analytical and numerical study of the fluorescence
spectra of a bistable driven system
by means of Keldysh diagram technique in pseudo--particle representation. The spectra exhibit smooth 
transition between ultra--quantum and the quasiclassical limits and 
indicate the threshold value of the external field when changing of the most probable stable state occurs.
The analysis of the fluorescence spectra also allows to determine the most probable stable state of the
system. It was also shown that at integer and half--integer detuning--nonlinearity ratio multiphoton resonance
leads to abrupt changes in fluorescence spectra.
It was also revealed that the fluorescence spectra are symmetric in the limit of zero environment 
temperature. The predicted features of the spectra could be observed in experiments with ultra-high quality 
resonators in either microwave or optical domain.
\end{abstract}
\maketitle

\section{Introduction}
Classical and quantum systems with two or more stable states
are under intense investigation for several decades.
Bistability and multistability often appear in the presence of an external coherent driving field \cite{Bonifacio1978} \cite{Gippius2007}. 
Discovered first in Fabry-Perot cavities 
filled with nonlinear medium \cite{Gibbs1976}, the phenomenon of bistability was observed in a wide range 
of experimental setups including laser cavities \cite{McCall1974}, whispering gallery resonators \cite{Braginsky1989}
exciton--polaritons in semiconductor microcavities \cite{Boulier2014}, etc. 
Recently, bistability at small photon numbers ($\sim 10^2$) has been observed in mesoscopic 
Josephson junction array resonators \cite{Muppalla2018} with ultra--high quality factor ($\sim 10^4$). 
Quantum optical and electronic systems exhibiting bistability are promising candidates for developing logic elements, switching and memory devices, various turnstiles, etc. 
Driven oscillator mode with Kerr--like 
nonlinearity is one of the simplest models describing bistability in various physical systems. 
The key features of bistable systems including S--shaped response curve and hysteresis cycle under varying external field amplitude can be analyzed by means of this model. 
This model also allows to obtain the stationary occupations and switching 
rates between two stable states due to interaction with 
dissipative environment \cite{Risken1987}, \cite{Vogel1990}, \cite{Maslova2019}, \cite{Maslova2019-1}.

The stationary populations of different stable states and switching rates in the model of driven nonlinear oscillator
in the presence of external noise have been thoroughly studied. The stationary density matrix
at zero environment temperature is obtained from the quantum kinetic equation of Fokker--Planck type for generalized Glauber P--function \cite{Drummond1980}. 
For small damping, numerical analysis based on rate equation was performed in
\cite{Risken1987}. The method suggested in \cite{Risken1987} allowed to calculate the switching rates and
the stationary occupations of the stable states for arbitrary temperature, which extended the analysis of \cite{Drummond1980}. The quasiclassical limit of the model was considered in \cite{Vogel1990}, 
where it was demonstrated that in the limit of large photon numbers and high temperature the model reduces to a classical stochastic one.
Such an approximation allowed to find analytical expressions for the stationary occupations of the quasienergy states and transition rates between two stable states for low damping.

Despite the well-studied statistical properties of the considered bistable system, 
there is a lack of detailed analysis of its fluorescence and transmitted light spectra.
In \cite{Drummond1980}, the spectrum was obtained by linearization of the Fokker--Planck--type kinetic equation for generalized Glauber function. 
A similar approach was exploited in \cite{Bonifacio1978}, where a mean-field model of atoms in resonant cavity was discussed. In \cite{Narducci1978} a factorization procedure of an infinite set of the equations for correlation functions
is used to get the spectrum. However, such procedures are valid only in the limit
of large damping constant and small quantum fluctuations. In the opposite limit of small damping, 
one should expect that the fluorescence spectrum is a combination of multiple narrow Lorenzian peaks corresponding to transitions between the quasienergy states, which totally differs from the quasi--Lorenzian linearized spectrum of \cite{Drummond1980}. 

The spectrum in this limit is not enough studied nowadays as well as in the intermediate regime where neither linearization could be performed nor the quasienergy states are well-defined. 
However, this regime is of particular interest in the context of bistability in Josephson junction array resonators \cite{Muppalla2018} because the nonlinearity per quantum in these resonators is comparable to the linewidth \cite{Weiss2015}. It is possible to find the spectrum in this regime by solving numerically the master equation introduced in \cite{Drummond1980}.
The details of spectrum behavior near the external field threshold value corresponding to the switching between the stable states are not clearly understood up to now. Moreover, the influence of multiphoton resonance leading to
degeneracy of quasienergy states \cite{Anikin2019} on fluorescence spectra also needs careful investigation.

In the present manuscript, we demonstrate that fluorescence
spectra of a bistable driven system indicate the external field threshold value and allows to find
the most probable state of the system. Our analysis is based on the calculation of polarization operator $\Pi^<$ by 
means of Keldysh diagram technique in pseudo--particle representation. In the general case, the equations for
the fluorescence spectrum are solved numerically, but analytical expressions are also presented in two limiting cases of small and large ratio between the detuning and damping constant. This allows to track the smooth transition between the classical and quantum fluorescence spectra behavior and to analyze the spectrum properties for nearly degenerate quasienergy states.

\section{The theoretical model}
We consider a simple model of bistable driven system consisting of a resonant mode with Kerr--like 
nonlinearity \cite{Drummond1980}, \cite{Risken1987}. The effective Hamiltonian of such system 
in the rotating--frame approximation reads
\begin{equation}
    \label{quant_bist_ham}
        \hat{H}_0 =-\Delta \hat{a}^\dagger \hat{a} + 
        \frac{\alpha}{2}(\hat{a}^\dagger \hat{a})^2 + f(\hat{a} + \hat{a}^\dagger).
\end{equation}
Here $\Delta$ is the detuning between the driving field and the resonant
oscillator frequency, $\alpha$ is the Kerr coefficient, and $f$ is proportional
to the amplitude of the driving field. 

\begin{figure}[h]
    \includegraphics[width=\linewidth]{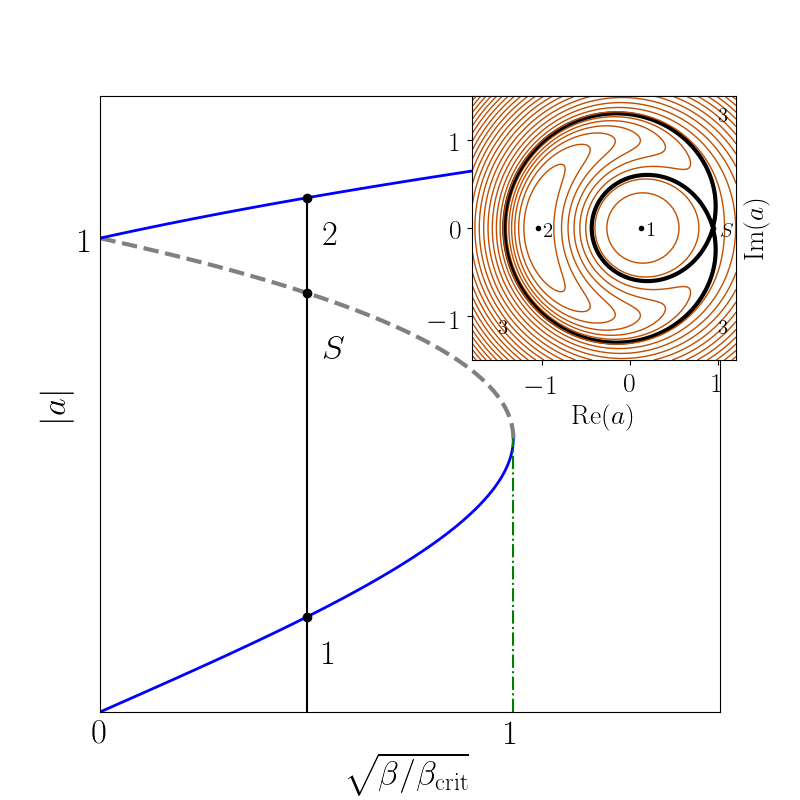}
    \caption{The S--shaped response curve to the external driving field 
             of the nonlinear oscillator model \eqref{quant_bist_ham} with
             $\Delta = \alpha = 1$ is shown. Blue solid lines denote the stable states 1 and 2, and gray dashed
             line corresponds to unstable state S. The classical phase portrait of the system 
             is shown on the inset for $\sqrt{\beta/\beta_\mathrm{crit}} = 0.3$, where 
             the stationary states are denoted by black dots.
             }
    \label{fig:phase_portrait}
\end{figure}

In the classical limit, one should replace the operators $\hat{a}$, $\hat{a}^\dagger$ 
in \eqref{quant_bist_ham} with classical field amplitudes 
$a, a^*$ to obtain the classical Hamiltonian. The S--shaped response curve is depicted in Fig.~\ref{fig:phase_portrait}.
On the inset, the classical phase portrait of the system is shown: the classical
trajectories in the $a$ plane are given by the contour lines of the classical Hamiltonian. On the phase
portrait, there are two stable stationary states 1 and 2, and one unstable stationary state S.
The only one dimensionless parameter of the classical Hamiltonian $\alpha f^2/\Delta^3 \equiv \beta$ governs the system dynamics. Bistability range is limited by maximal value of $\beta = \beta_\mathrm{crit} \equiv 4/27$.
For the quantum Hamiltonian, there exists another dimensionless parameter $m \equiv 2\Delta/\alpha$. The quasiclassical limit is acquired at large non--integer values of $m$.

Let us assume that the system is weakly interacting with the environment, so the full Hamiltonian reads
\begin{equation}
    \hat{H}_\mathrm{full} = \hat{H}_0 + \hat{\xi}^\dagger \hat{a} + \hat{\xi} \hat{a}^\dagger.
\end{equation}
For the case of white noise, the damping operators $\hat{\xi}, \hat{\xi}^\dagger$ are delta--correlated:
\begin{equation}
    \label{noise_correlators}
    \begin{gathered}
        \avg{\hat{\xi}(t)\hat{\xi}^\dagger(t')} = \gamma(N + 1)\delta(t-t'),\\
        \avg{\hat{\xi}^\dagger(t)\hat{\xi}(t')} = \gamma N\delta(t-t').\\
    \end{gathered}
\end{equation}
Here $N$ is a number of thermal photons at the external field frequency.

With such assumptions, the evolution of the system density matrix can be described by the master equation:
\cite{Haken1965}, \cite{Risken1965},
\cite{Graham1970}, \cite{Drummond1980}, \cite{Risken1987}:
\begin{multline}
    \label{lindblad_eq}
    \p_t \hat{\rho} = \mathcal{L}\hat{\rho} = -i[\hat{H}_0, \hat{\rho}] + \frac{\gamma}{2}
                    \left(2\hat{a}\hat{\rho} \hat{a}^\dagger - 
                \hat{\rho} \hat{a}^\dagger \hat{a}\right.\\ - \hat{a}^\dagger \hat{a} \hat{\rho}
                                     \left. + 2N[[\hat{a},\hat{\rho}],\hat{a}^\dagger]\right).
\end{multline}
This equation can be derived in quasienergy representation 
Equation for $\hat{\rho}$ follows from kinetic equation for lesser $G^<_{nn'}(t,t)$ 
obtained by means of Keldysh diagram technique generalized for pseudo--particle approach with 
additional constraint on physically available states (see Appendix \ref{appendix:A}), 
where $|n\rangle$ are eigenstates of the system Hamiltonian $\hat{H}_0$.

The incoherent part of the photoluminescence spectrum is given by the correlation function of the 
operators $\hat{a}, \hat{a}^\dagger$ \cite{Drummond1980}:
\begin{equation}
    S(\omega) = \int dt e^{i\omega t} \avg{\hat{a}^\dagger(0) \hat{a}(t)}.
\end{equation}
To obtain $S(\omega)$, one should calculate the polarization operator $\Pi^<(\omega)$ in pseudo--particle 
Keldysh diagram technique (see Appendix \ref{appendix:B}), which coincides with $S(\omega)$ in stationary case.

The detailed calculation of $\Pi^<$ given in the Appendix \ref{appendix:B}
leads to the following expression for fluorescence spectrum:
\begin{equation}
    \label{exact_spectrum}
    S(\omega) = 2\Re\Tr{\left\{\hat{a} [-i\omega\check{\mathds{1}} - 
                    \check{\mathcal{L}}]^{-1} \hat{\rho}_\mathrm{st}\hat{a}^\dagger\right\}}.
\end{equation}
Here the superoperator $\mathcal{L}$ is determined by Eq.~\eqref{lindblad_superoperator} and has 
the following explicit form for white noise:
\begin{multline}
    \mathcal{L}_{ij,kl} = -i(\epsilon_i - \epsilon_j)\delta_{ik}\delta_{lj}\\ + 
            \frac{\gamma(N+1)}{2}
            \left(2a_{ik}a^*_{jl} - (\hat{a}^\dagger \hat{a})_{ik}\delta_{lj} - 
                \delta_{ik}(\hat{a}^\dagger \hat{a})_{lj}\right)\\
            +\frac{\gamma N}{2}
            \left(2a^*_{ki}a_{lj} - (\hat{a}\hat{a}^\dagger)_{ik}\delta_{lj} - 
                                    \delta_{ik}(\hat{a}\hat{a}^\dagger)_{lj}\right).
\end{multline}
The stationary density matrix can obtained from kinetic equation \eqref{lindblad_eq}.
The incoherent part of the spectra can be calculated numerically by means of 
Eq.~\eqref{exact_spectrum}.

There exist two complementary approximations allowing to get analytical expressions for $S(\omega)$. 
They correspond to the quantum limit of well-defined quasienergy states and the 
quasiclassical limit of small quantum and thermal fluctuations considered in \cite{Drummond1980}.

The first approximation is valid providing that
$\gamma$ is smaller than any of the differences between eigenvalues of the effective Hamiltonian, the spectrum
consists of distinct Lorenzian peaks corresponding to transitions between quasienergy levels. When $\omega$ is
close to a transition frequency between the levels $|n\rangle$ and $|n'\rangle$, 
the following expression for the spectrum is valid (see Appendix \ref{appendix:B})
\begin{equation}
    \label{lorentz_spectrum}
    S(\omega) = 
            \sum_{n'n}P_{n'} |a_{nn'}|^2 \frac{2\Gamma_{n'n}}
            {(\omega - \epsilon_n + \epsilon_{n'})^2 + \Gamma_{n'n}^2},
\end{equation}
where the widths of the Lorenzian peaks are defined as
\begin{multline}
    2\Gamma_{n'n} = 
        \gamma(N+1)\left((\hat{a}^\dagger \hat{a})_{n'n'} + 
            (\hat{a}^\dagger \hat{a})_{nn} -2\Re{a_{n'n'}a_{nn}^*}\right)\\ +
        \gamma N\left((\hat{a}\hat{a}^\dagger)_{n'n'} + (\hat{a} \hat{a}^\dagger)_{nn} 
        -2\Re{a_{n'n'}^*a_{nn}} \right),
\end{multline}
and $P_{n'}$ are the stationary occupation probabilities of the unperturbed Hamiltonian 
eigenstates $|n'\rangle$. 
The occupation probabilities in this approximation are defined by the stationary rate 
equation \cite{Risken1987} which is the diagonal approximation of the quantum master equation \eqref{lindblad_eq}.

Another approximation exploited in \cite{Drummond1980} is based on quantum Fokker--Planck equation for 
density matrix in generalized P representation. 
\begin{multline}
    \frac{\p P(a,a^*)}{\p t} = 
    \left[i\frac{\p}{\p a}((-\Delta -i\gamma/2) a + \alpha a^2 a^* + f)\right.\\
     - \left.\frac{i\alpha}{2}\frac{\p^2}{\p a^2} a^2
    + \frac{\gamma N}{2} \frac{\p^2}{\p a \p a^*}\right]P(a,a^*) + \mathrm{c.c.}
\end{multline}
This equation can be linearized near each of the classical stable states 
which can be obtained from the solution of stationary classical equations of motion:
\begin{equation}
    -(\Delta +i\gamma/2) a + \alpha a^2 a^* + f = 0.
\end{equation}
The resulting expression for spectrum reads
\begin{multline}
    \label{drummond_walls_spectrum}
    S(\omega) = \sum_{q=1,2}\frac{1}{4\pi^2|\lambda_q(\omega)|^2}
    \gamma\left[(1 + N)\alpha^2 n_q^2 \right.\\
        \left. +N|\omega - \Delta + 2\alpha n_q - i\gamma/2|^2\right] P_q
\end{multline}
with
\begin{equation}
    \lambda_q(\omega) = -(\omega - \Delta + 2\alpha n_q - i\gamma/2)
              (\omega + \Delta - 2\alpha n_q - i\gamma/2) - \alpha^2 n_q^2,
\end{equation}
where $q = 1,2$ correspond to two stationary states, $P_q$ are classical probabilities to find the system
near each of the stable states $q$, and $n_q$ is the mean value of $\hat{a}^\dagger \hat{a}$ in the stable state $q$. 

\section{Results and discussion}
Using the approach described in the previous section, we calculated the incoherent part of the emission spectrum
of the quantum driven nonlinear oscillator. We considered different values of the model parameters including the small--fluctuation regime and weak--coupling regime.

\begin{figure}[h!]
    \centering
    \includegraphics[width=\linewidth]{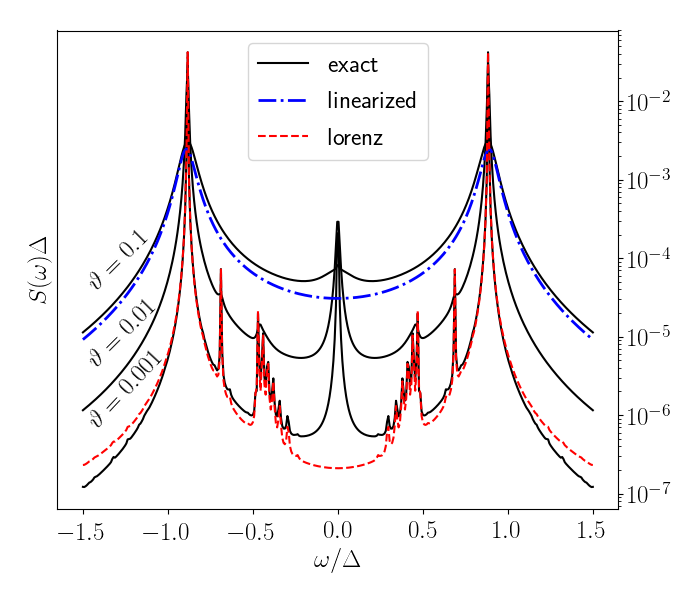}
    \caption{
        The fluorescence spectra of the quantum driven nonlinear oscillator model 
        are shown for $m = 12.5$, $N = 0$, $f/f_\mathrm{crit} = 0.2$ and 
        different values of $\vartheta$. Black solid lines denote the spectra obtained 
        from Eq.~\eqref{exact_spectrum}, red dashed line represents the spectrum obtained from Lorenzian 
        approximation, and blue dashdotted line represents the spectrum obtained from linear approximation.
        At small $\vartheta$, the spectrum consists of multiple narrow Lorenzian peaks
        corresponding to transitions between different quasienergy levels, and it is well 
        approximated by Eq.~\eqref{lorentz_spectrum}.
        With increasing $\vartheta$, the spectrum smoothly transforms to a set of quasi--Lorenzian peaks 
        corresponding to linearized approximation determined by Eq.~\eqref{drummond_walls_spectrum}.
        }
    \label{fig:theta}
\end{figure}

First of all, we explored the transition between the small--fluctuation
and weak--coupling regimes which occurs at decreasing dimensionless damping constant 
$\vartheta \equiv \gamma/\Delta$ (see Fig.~\ref{fig:theta}). 
At small $\vartheta$, the exact spectrum matches with 
the Lorenzian approximation \eqref{lorentz_spectrum} and consists of many distinct peaks corresponding to transitions
between the quasienergy states.
With increasing $\vartheta$, it becomes necessary to solve the kinetic equations
taking into account all non--diagonal elements of the density matrix.
For rather large $\vartheta$, the solution of the master equation leads to the same result as
the linear approximation of Eq.~\eqref{drummond_walls_spectrum} corresponding to the vicinity of each classical
stable state. The relative height of the peaks is determined by the occupation probabilities of these stable states.
The spectrum defined by Eq.~\eqref{drummond_walls_spectrum} consists of two quasi--Lorenzian
peaks located at $\pm \omega_{1,2}$, where $\omega_{1,2}$ are the frequencies of the oscillator motion around the stable states 1 and 2. It is important to mention that a peak at zero frequency is always present in the fluorescence spectrum. This peak is connected with transitions between the classical stable states lying in different
regions of the system phase portrait.

\begin{figure}[h!]
    \centering
        \centering
        \includegraphics[width=\linewidth]{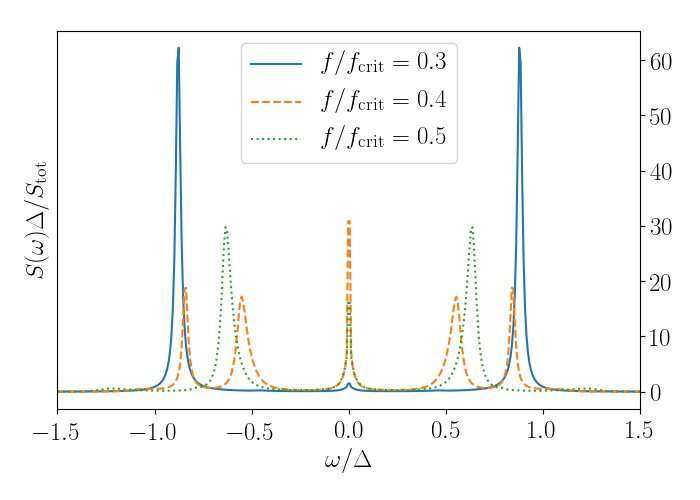}
        \caption{The fluorescence spectra  $S(\omega)$ 
            of the quantum driven nonlinear oscillator are shown for 
            $m = 12.5$, $\vartheta = 0.03$, $N = 0$ and varying values of $f$. The spectra are 
            normalized by total fluorescence intensity 
            $S_\mathrm{tot} \equiv \int \frac{d\omega}{2\pi}S(\omega)$.
            For the external fields below threshold ($f/f_\mathrm{crit} = 0.3$) and 
            above threshold ($f/f_\mathrm{crit} = 0.5$), 
            two symmetric side peaks at nonzero frequencies corresponding to different
            classical stable states are present. 
            Near the threshold, the spectrum contains four side peaks corresponding 
            to both classical stable states.
            }
        \label{fig:f}
\end{figure}

Another interesting feature is that both in weak--coupling and small--fluctuation regimes the spectra strongly depend on the value of the external driving field, which is demonstrated in Fig.~\ref{fig:f}. In particular, in the small--fluctuation 
regime when the Eq.~\eqref{drummond_walls_spectrum} is applicable, the spectrum explicitly depends on $n$. 
Obtained within the linearized
approximation in the vicinity of each stable state the spectrum defined by 
Eq.~\eqref{drummond_walls_spectrum} contains two peaks at nonzero frequency. However, the position of these peaks
depends on $n$ and therefore
differs for the driving fields below threshold ($f < f_0$) and above threshold ($f > f_0$). At the threshold value
of external field $f_0$, switching between the most probable states $1$ and $2$ takes place.
In the vicinity of the threshold, $f \approx f_0$, the system can be found in both stable states with comparable probabilities, and the total spectrum contains contributions from both of them. Therefore, the spectrum exhibits a crossover when the external field passes the threshold value. Below the threshold, it contains two peaks at frequencies
$\pm \omega_1$ corresponding to the stable state $1$, 
in the vicinity of the threshold, it contains four peaks at frequencies $\pm\omega_1, \pm\omega_2$ corresponding to both stable states, and above the threshold, there are two peaks again at frequencies $\pm\omega_2$ corresponding to.
stable state $2$.
Similar arguments can be applied to the weak--coupling regime. The occupation numbers of the eigenstates also differ drastically for 
$f < f_0$ and $f > f_0$, and different quasienergy states contribute to the spectrum in these cases. 

\begin{figure}[h!]
    \centering
    \includegraphics[width=\linewidth]{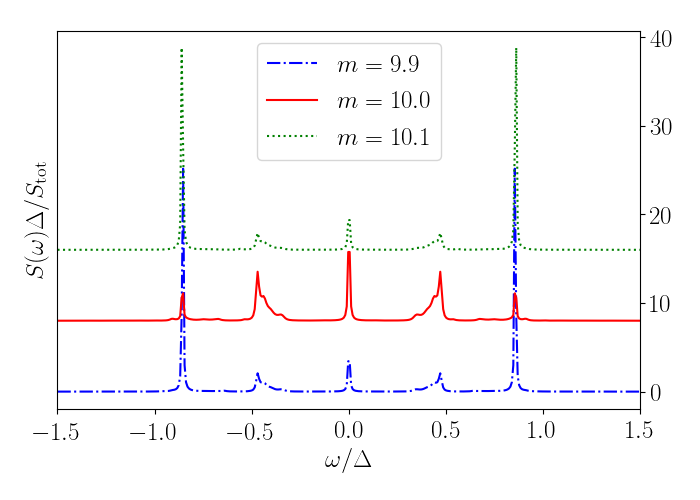}
    \caption{The fluorescence spectra of the quantum driven nonlinear oscillator are shown for 
        $f/f_\mathrm{crit} = 0.29$, 
        $\vartheta = 0.005$, $N = 0$ and varying values of $m$. For clarity, different spectra are shifted 
        vertically by arbitrary value. With varying $m$, the relative heights of 
        the peaks corresponding to different classical stable states strongly change.}
    \label{fig:m}
\end{figure}

Also, we considered the behavior of spectra near the integer values of the ratio between double detuning and 
nonlinearity $m$ (see Fig.~\ref{fig:m}). 
As shown in \cite{Anikin2019}, at integer values of $m$ the multiphoton resonance leads to the enhanced probability of the system to occupy the classical stable state with higher amplitude. This could be interpreted as a decrease in the 
threshold value $f_0$ of the driving field near the integer values of $m$. 
So, varying $m$ changes the relative heights of the peaks corresponding to different classical stable states. 
When $m$ is
close to an integer, the height of the spectrum
peaks corresponding to the stable state 2 abruptly increases whereas the magnitude of the peaks 
corresponding to the stable state 1 strongly decreases.

\begin{figure}[h!]
    \centering
    \includegraphics[width=\linewidth]{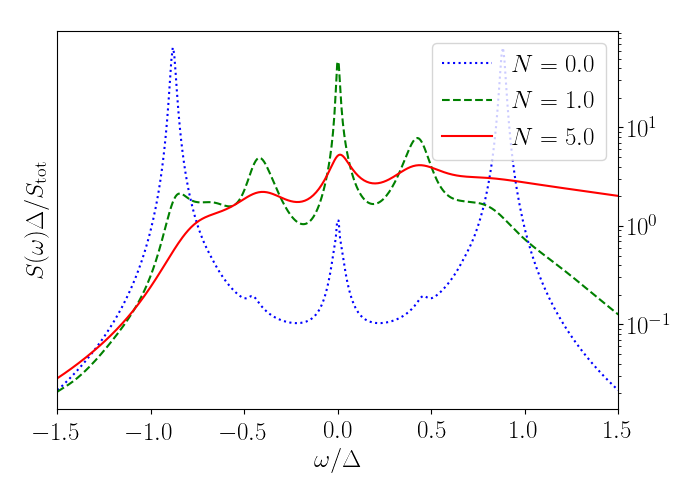}
    \caption{The fluorescence spectra are depicted for $f/f_\mathrm{crit} = 0.29$, 
                $\vartheta = 0.03$, $m = 12.5$ and varying $N$. Mirror symmetry with respect to 
            vertical axis is present at $N = 0$, and it is violated at $N \ne 0$.}
    \label{fig:nth}
\end{figure}

In addition, we considered the spectra at different bath temperatures. As can be seen on Fig.~\ref{fig:nth}, 
the spectra are perfectly symmetric at $N = 0$ and becomes asymmetric at $N > 0$.


\section{Conclusions}
The fluorescence spectra of a bistable driven system were studied analytically and numerically by means of Keldysh
diagram technique in pseudo--particle representation. The spectra exhibit a smooth transition between ultra--quantum and the quasiclassical limits.
It was shown that fluorescence spectra indicate the external field threshold value which corresponds to switching between the most probable states of a bistable system. Moreover,
in the vicinity of the external field threshold value the system can be found with comparable probabilities in both stable
states. Thus, nearly equal contributions from both states can be clearly seen in the total spectrum. 
So, it is possible to determine the most probable state of the system from fluorescence spectra in a 
wide range of external field intensities. In addition, it was revealed that at integer and half--integer values of the detuning--nonlinearity ratio multiphoton resonance leads to an enhanced probability of the system to occupy the classical stable state with higher amplitude.
So, when this ratio is close to an integer or half--integer, the height of the spectrum peaks corresponding
to the stable state 2 with higher amplitude abruptly increases whereas the magnitude of the peaks
corresponding to the stable state 1 with lower amplitude strongly decreases. 
Also we found out that the fluorescence spectra are symmetric at zero bath temperature, and this symmetry breaks down with increasing temperature.

\begin{acknowledgements}
    This work was supported by RFBR grants 19--02--000--87a, 18--29--20032mk, 19-32-90169 and by a grant of the Foundation for the Advancement of Theoretical Physics and Mathematics ''Basis''.
\end{acknowledgements}

%

\appendix
\section{The derivation of master equation}
\label{appendix:A}

The Hamiltonian $\hat{H}_0$ can be diagonalized so that $|n\rangle$ is the set of its eigenvectors.
It is convenient to introduce pseudo--particle operators $\hat{c}_n^\dagger, \hat{c}_n$ which correspond to 
creation/annihilation
of the system eigenstates $|n\rangle$. This implies the constraint on the space of possible physical states:
$\sum_n \hat{c}_n^\dagger \hat{c}_n = \mathds{1}$.
The operators $\hat{H}_0$, $\hat{a}$, $\hat{a}^\dagger$ can be expressed through 
pseudo--particle creation/annihilation operators $\hat{c}^\dagger_n, \hat{c}$:
\begin{equation}
    \begin{gathered}
        \hat{H}_0 = \sum_n\epsilon_{n} \hat{c}_n^\dagger \hat{c}_n\\
        \hat{a} = \sum_{nn'} a_{nn'} \hat{c}_n^\dagger \hat{c}_{n'}.\\
    \end{gathered}
\end{equation}

Although it does not affect physical results, we assume that the pseudo--particles are fermions.
We will derive the kinetic equation for the density matrix of the system which is
directly related to the $G^<(t,t')$ Green function:
$\rho_{nn'}(t) = -iG^<_{nn'}(t,t)$. 

The constraint on the Hilbert space, which means the presence of exactly one pseudoparticle in the system,
leads to the following additional rules for constructing Keldysh diagram technique \cite{Arseev2014},
\cite{Arseev2015}:
\begin{enumerate}
    \item Among all diagrams only the diagrams with one $G^{<}_0$ line are kept,
    which denotes a one--particle Green function of non--interacting system
    \item If $G^<$ is present in some diagram, the other Green functions cannot contain the pseudo--particle 
    occupation numbers. Therefore, in all diagrams containing $G^<$ Green functions
    $G^>$ reduces to $G^R - G^A$, and the self--energy parts for $G^{R,A}$ 
    also don't contain $G^<$.
\end{enumerate}

\begin{figure}[h]
    \centering
    \vspace{-0.3cm}
    \begin{minipage}[c]{\linewidth}
    \begin{equation*}
        \begin{tikzpicture}[baseline={([yshift=-0.1cm]current bounding box.center)}]
            \begin{feynman}
                \vertex  (a);
                \vertex[right=1.1cm of a] (b) {};
            
                \diagram* {
                    (a) -- [fermion, ultra thick] (b)
                };
            \end{feynman}
        \end{tikzpicture}
        =\hspace{0.2cm}
        \begin{tikzpicture}[baseline={([yshift=-0.1cm]current bounding box.center)}]
            \begin{feynman}
                \vertex  (a);
                \vertex[right=1.1cm of a] (b) {};
            
                \diagram* {
                    (a) -- [fermion] (b)
                };
            \end{feynman}
        \end{tikzpicture}
        +\hspace{0.55em}
        \begin{tikzpicture}[baseline={([yshift=-0.30cm]current bounding box.center)}]
            \begin{feynman}
                \vertex  (a);
                \vertex[right=1.1cm of a] (b);
                \vertex[right=1.1cm of b] (c);
                \vertex[right=1.1cm of c] (d);
                \vertex[above right=0.55cm and 0.55cm of b] (e);

                \diagram* {
                    (a) -- [fermion] (b) -- [fermion, ultra thick] (c) -- [fermion, ultra thick] (d),
                    (b) -- [scalar, quarter left] (e) -- [scalar, quarter left] (c)
                };
            \end{feynman}
        \end{tikzpicture}
    \end{equation*}
\end{minipage}
    \caption{
            Graphical representation of Dyson equation for $G^R$, $G^A$, $G^<$ is shown.
            Thin solid line corresponds to pseudo--particle Green function, thick solid line 
            corresponds to dressed pseudo--particle
            Green function, dashed line corresponds to bath correlation function.
            }
    \label{fig:dyson}
\end{figure}
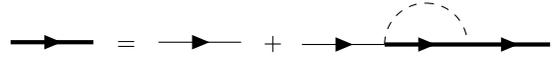
In self--consistent Born approximation, diagrams for $G^R$, $G^A$, $G^<$ are shown in Fig.~\ref{fig:dyson}.
In the case of white noise, the diagrams with crossing bath Green functions 
(shown by dashed lines on the Fig.~\ref{fig:dyson}) 
vanish which provides the validity of Born approximation.
Sum of all diagrams depicted on Fig.~\ref{fig:dyson} can be represented by self--consistent equations
\begin{equation}
    \label{green_functions}
    \begin{gathered}
       G^< = (1 + G^R\Sigma^R)G^<_0(1 + \Sigma^AG^A) + G^R\Sigma^<G^A,\\
        G^R = G_0^R + G_0^R \Sigma^R G^R,\\
        G^A = G_0^A + G^A \Sigma^A G_0^A,\\
    \end{gathered}
\end{equation}
where 
$(G^{R,A}_0)_{nn'}(t,t') = \mp ie^{-i\epsilon_n(t-t')}\theta(\pm(t-t'))\delta_{nn'}$, 
$(G^<_{0})_{nn'}(t,t') = ie^{-i\epsilon_n(t-t')}n_n\delta_{nn'}$, $n_n$ are pseudoparticle occupation numbers,
$\Sigma^{R,A}$ are sums of irreducibble diagrams containing no $G^<_0$ lines, and $\Sigma^<$ is the sum 
of irreducible diagrams containing exactly one $G^<_0$ line. In self--consistent approximation, they can
be expressed as
\begin{multline}
        \Sigma^{R(A)}_{ij}(t,t') = i\sum_{kl}a_{ik}G^{R(A)}_{kl}(t,t') D^{>(<)}(t,t') a^*_{jl} \\+
                        a^*_{ki}G^{R(A)}_{kl}(t,t') D^{<(>)}(t',t) a_{lj},\\
\end{multline}
\begin{multline}
    \Sigma^<_{ij}(t,t') = i\sum_{kl}a_{ik} G^<_{kl}(t,t') D^>(t,t') a^*_{jl}\\ +
                 i\sum_{kl}a^*_{ki} G^<_{kl}(t,t') D^<(t,t') a_{lj},
\end{multline}
where
\begin{equation}
    \label{d_green_functions}
    \begin{gathered}
        D^<(t,t') = -i\avg{\hat{\xi}^\dagger(t') \hat{\xi}(t)},\\
        D^>(t,t') = -i\avg{\hat{\xi}(t)\hat{\xi}^\dagger(t') },
    \end{gathered}
\end{equation}
and the correlation functions of $\hat{\xi},\hat{\xi}^\dagger$ are defined by Eq.~\eqref{noise_correlators}.
By applying $G_0^{-1} \equiv i\p_t - \hat{H}_0$ to the first 
equation \eqref{green_functions} from the left and from the right, one gets the following equation:
\begin{multline}
    \label{quant_kin_eq}
    (i\p_t + i\p_{t'} - \epsilon_n + \epsilon_{n'})G^<_{nn'}(t,t') = \\ \int dt''\,
    \left(\Sigma^R(t,t'') G^<(t'',t')  + \Sigma^<(t,t'') G^A(t'',t') \right. \\
       \left. -G^R(t,t'')\Sigma^<(t'',t') -  G^<(t,t'') \Sigma^A(t'',t')\right)_{nn'},
\end{multline}
where self--energies $\Sigma^R$, $\Sigma^A$, $\Sigma^<$ are defined as
\begin{equation}
    \Sigma^{R(A)}_{n_1n_2}(t,t') = \frac{i\gamma\delta(t-t')}{2}\left[
        (1+N)(\hat{a}^\dagger \hat{a})_{n_1n_2} + N(\hat{a}\hat{a}^\dagger)_{n_1n_2}\right],
\end{equation}
\begin{multline}
       \Sigma^<_{n_1n_2} = \gamma \delta(t-t')\left[(1 + N) a_{n_1k} G^<_{kk'}(t,t) a^*_{n_2k'}\right.\\+
            \left. N a^*_{kn_1} G^<_{kk'}(t,t) a_{k'n_2}\right].
\end{multline}
In the above equation and below, we omit the summation over repeating indices.
After substituting these expressions into Eq.~\eqref{quant_kin_eq}, we obtained exactly the quantum master equation
for delta--correlated bath:
\begin{multline}
    \label{lindblad_eq_ind}
    (i\p_t - \epsilon_i + \epsilon_j)G^<_{nn'}(t,t) = \\
        \frac{i\gamma (1+ N)}{2} \left(2a_{nk} a^*_{n'l} - 
        (\hat{a}^\dagger \hat{a})_{nk}\delta_{n'l} - 
        \delta_{nk}(\hat{a}^\dagger \hat{a})_{n'l}\right) G^<_{kl}(t,t)\\ + 
        \frac{i\gamma N}{2} \left(2a^*_{kn} a_{ln'} - 
        (\hat{a}\hat{a}^\dagger)_{nk}\delta_{n'l} - \delta_{nk}(\hat{a}\hat{a}^\dagger)_{n'l}\right) G^<_{kl}(t,t).
\end{multline}
The equation \eqref{lindblad_eq_ind} is equivalent to the quantum master equation Eq.~\eqref{lindblad_eq}:
\begin{multline}
    \label{lindblad_eq_1}
    \p_t \rho_{nn'} = (\mathcal{L}\rho)_{nn'} = -i(\epsilon_n - \epsilon_{n'})\rho_{nn'} \\+ \frac{\gamma}{2}
                    \left(2a\rho a^\dagger - \rho a^\dagger a\right. - a^\dagger a \rho 
                                     \left. + 2N[[a,\rho],a^\dagger]\right)_{nn'}.
\end{multline}

\section{Polarization operator}
\label{appendix:B}
\begin{figure}[h]
    \centering
    \begin{minipage}[c]{\linewidth}
    \begin{equation*}
        \Pi^< = 
        \begin{tikzpicture}[baseline={([yshift=-0.1cm]current bounding box.center)}]
            \begin{feynman}
                \vertex [label={left:$\hat{a}$}] (a);
                \vertex[right=2.5cm of a,draw, circle] (b) {$\Phi_1$};
            
                \diagram* {
                    (a) -- [fermion, quarter left, edge label=$<$] (b),
                    (b) -- [fermion, quarter left, edge label=$R$] (a),
                };
            \end{feynman}
        \end{tikzpicture}
        \quad-\quad
        \begin{tikzpicture}[baseline={([yshift=-0.1cm]current bounding box.center)}]
            \begin{feynman}
                \vertex[draw, circle] (a) {$\Phi_2$};
                \vertex[right=2.75cm of a, label={right:$\hat{a}^\dagger$}] (b);
            
                \diagram* {
                    (a) -- [fermion, quarter left, edge label=$<$] (b),
                    (b) -- [fermion, quarter left, edge label=$A$] (a),
                };
            \end{feynman}
        \end{tikzpicture}
    \end{equation*}

    \begin{equation*}
        \begin{tikzpicture}[baseline={([yshift=-0.1cm]current bounding box.center)}]
            \begin{feynman}
                \vertex (a) at (-0.6, 0.6);
                \vertex[draw, circle] (phi)  at (0,0) {$\Phi_1$};
                \vertex (b) at (-0.6, -0.6);
                
                \diagram* {
                    (a) -- (phi) -- (b)
                };
            \end{feynman}
        \end{tikzpicture} 
        \hspace{0.5em}
        =
        \hspace{0.5em}
        \begin{tikzpicture}[baseline={([yshift=-0.1cm]current bounding box.center)}]
            \begin{feynman}
                \vertex (a) at (-0.5, 0.5);
                \vertex[label={right:$\hat{a}^\dagger$}] (op)  at (0,0);
                \vertex (b) at (-0.5, -0.5);
                
                \diagram* {
                    (a) -- (op) -- (b)
                };
            \end{feynman}
        \end{tikzpicture} 
        \hspace{0.5em}
        +
        \hspace{0.5em}
        \begin{tikzpicture}[baseline={([yshift=-0.1cm]current bounding box.center)}]
            \begin{feynman}
                \vertex (a) at (0,0.7);
                \vertex (b) at (0.5,0.7);
                \vertex[draw, circle] (phi) at (1.5,0) {$\Phi_1$} ;
                \vertex (c) at (0.5, -0.7);
                \vertex (d) at (0, -0.7);
                
                \diagram* {
                    (a) --  (b) -- [fermion, bend left=20, edge label={$A$}] (phi) -- [fermion, bend left=20, edge label = {$R$}] (c) --  (d),
                    (b) -- [scalar] (c),
                };
            \end{feynman}
        \end{tikzpicture} 
    \end{equation*}

    \begin{equation*}
        \begin{tikzpicture}[baseline={([yshift=-0.1cm]current bounding box.center)}]
            \begin{feynman}
                \vertex (a) at (0.6, 0.6);
                \vertex[draw, circle] (phi)  at (0,0) {$\Phi_2$};
                \vertex (b) at (0.6, -0.6);
                
                \diagram* {
                    (a) -- (phi) -- (b)
                };
            \end{feynman}
        \end{tikzpicture} 
        \hspace{0.5em}
        =
        \hspace{0.5em}
        \begin{tikzpicture}[baseline={([yshift=-0.1cm]current bounding box.center)}]
            \begin{feynman}
                \vertex (a) at (0.5, 0.5);
                \vertex[label={left:$\hat{a}$}] (op)  at (0,0);
                \vertex (b) at (0.5, -0.5);
                
                \diagram* {
                    (a) -- (op) -- (b)
                };
            \end{feynman}
        \end{tikzpicture} 
        \hspace{0.5em}
        +
        \hspace{0.5em}
        \begin{tikzpicture}[baseline={([yshift=-0.1cm]current bounding box.center)}]
            \begin{feynman}
                \vertex (a) at (0,0.7);
                \vertex (b) at (-0.5,0.7);
                \vertex[draw, circle] (phi) at (-1.5,0) {$\Phi_2$} ;
                \vertex (c) at (-0.5, -0.7);
                \vertex (d) at (0, -0.7);
                
                \diagram* {
                    (d) --  (c) -- [fermion, bend left=20, edge label={$A$}] (phi) -- [fermion, bend left=20, edge label={$R$}] (b) --  (a),
                    (b) -- [scalar] (c),
                };
            \end{feynman}
        \end{tikzpicture} 
    \end{equation*}
\end{minipage}
    \caption{Diagram ladder representing the polarization operator $\Pi^<$ is shown. Solid lines represent 
            pseudo--particle fermion Green functions, dashed lines represent bath Green functions, and 
            circles denote effective vertices $\Phi_{1,2}$.} 
    \label{fig:polarization_operator}
\end{figure}
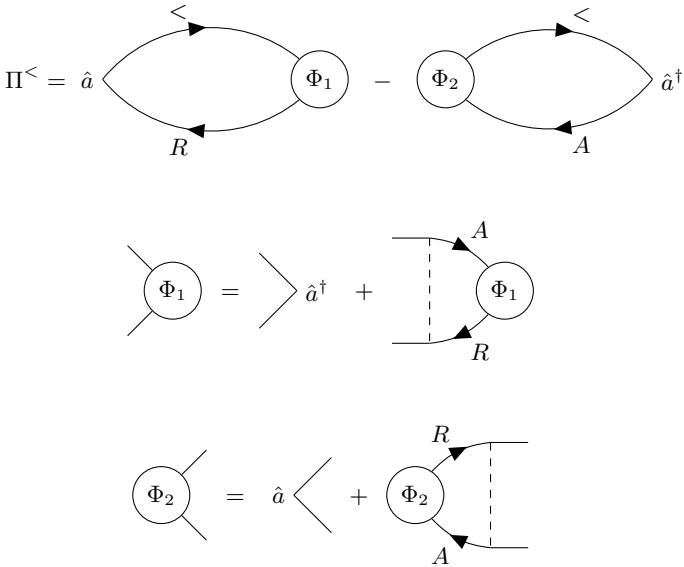

The polarization operator 
$\Pi^<(t,t') \equiv \avg{\mathrm{T}_C \hat{a}_-(t) \hat{a}_+^\dagger(t')}$ 
is represented as a 
sum of diagrams shown on Fig.~\ref{fig:polarization_operator}. 
For white external noise, it consists of a ladder of diagrams without crossing
bath propagators. Sum of all such diagrams can be calculated in a way similar to Bethe--Salpether equation.
In steady state, $\Pi^<(t,t')$ depends only on $t-t'$. Thus, in $\omega$--representation the graphical equations
shown on Fig.~\ref{fig:polarization_operator} take form
\begin{multline}
    \label{polarization_operator}
    \Pi^<(\omega) = 
            \int \frac{d\omega'}{2\pi} \Tr \hat{a} G^<(\omega+\omega') \Phi_1(\omega) G^R(\omega')\\
            -\int \frac{d\omega'}{2\pi} \Tr \Phi_2(\omega)G^<(\omega'+\omega) \hat{a}^\dagger G^A(\omega'),
\end{multline}
where $G^R(\omega)$, $G^A(\omega)$ are the solutions of Eq.~\eqref{green_functions} in Fourier representation.
Stationary green function $G^<(\omega)$ obtained from these equations takes form 
$G^<(\omega)_{ij} = (\rho_\mathrm{st}G^A(\omega))_{ij} - (G^R(\omega)\rho_\mathrm{st})_{ij}$, 
$(\rho_\mathrm{st})_{ij} \equiv -i\int \frac{d\omega}{2\pi} G^<(\omega)_{ij}$.
The vertex functions $\Phi_1(\omega)$, $\Phi_2(\omega)$ read
\begin{multline}
    \label{phi_1_dyson}
    \Phi_{1}(\omega)_{kl} = a^*_{lk}\\
               +\gamma\int \frac{d\omega'}{2\pi}(1+N) 
                (\hat{a} G^A(\omega+\omega') \Phi_1(\omega) G^R(\omega') \hat{a}^\dagger)_{kl}\\ + 
                            N (\hat{a}^\dagger G^A(\omega+\omega') \Phi_1(\omega) G^R(\omega') \hat{a})_{kl},
\end{multline}
\begin{multline}
    \label{phi_2_dyson}
    \Phi_2(\omega)_{kl} =  a_{kl}\\ + \gamma\int \frac{d\omega'}{2\pi}(1+N) 
            (\ha G^A(\omega') \Phi_2(\omega) G^R(\omega'+\omega) \ha^\dagger)_{kl}\\+ 
                            N (\ha^\dagger G^A(\omega')\Phi_2(\omega) G^R(\omega'+\omega) \ha)_{kl}.
\end{multline}
From \eqref{phi_1_dyson} and \eqref{phi_2_dyson}, one could obtain $\Phi_{1,2}(\omega)$ explicitly. 
For this, it is convenient to introduce tensor notation for denoting superoperators acting on operators, because 
the space of operators acting on Hilbert space is isomorphic to the tensor product of a Hilbert space and its
conjugate. For example, the combination $\ha G^A(\omega') \Phi_2(\omega) G^R(\omega+\omega') \ha^\dagger$ 
can be written using this notation
as $\left(\ha G^A(\omega') \otimes (G^R(\omega+\omega')\ha^\dagger)^T\right) \Phi_2(\omega)$: this means 
that the operator
$\ha G^A(\omega') \otimes (G^R(\omega+\omega')\ha^\dagger)^T$ acting on the tensor product of a Hilbert space and its
conjugate is multiplied on $\Phi_2$ which is considered as a vector from the tensor product. 

Using this notations, one gets the expressions for $\Phi_{1,2}$:
\begin{multline}
    \label{gamma_1}
    \Phi_1(\omega) = \left[\mathds{1}\otimes\mathds{1} - 
        \gamma\int\frac{d\omega'}{2\pi}
        \left[(1+N) \ha^\dagger \otimes \ha^T \right.\right.\\\left.\left.+ N \ha \otimes (\ha^\dagger)^T\right] G^A(\omega+\omega')\otimes G^R(\omega')^T
                        \vphantom{\int}\right]^{-1}\ha^\dagger,
\end{multline}
\begin{multline}
    \label{gamma_2}
    \Phi_2(\omega) = \left[\mathds{1}\otimes\mathds{1} - 
        \gamma\int\frac{d\omega'}{2\pi}
        \left[(1+N) \ha^\dagger\otimes \ha^T \right.\right.\\
        \left.\left.+ N \ha \otimes (\ha^\dagger)^T\right] G^A(\omega')
                \otimes G^R(\omega+\omega')^T \vphantom{\int}\right]^{-1}\ha.
\end{multline}
Using the equations \eqref{polarization_operator}, \eqref{gamma_1}, \eqref{gamma_2}, it is possible to calculate
polarization operator in a closed form. Moreover, all these equations contain
integral of product of retarded and advanced Green functions. For this integral, the following identity holds:
\begin{multline}
    \label{GRGA_integral}
    \int \frac{d\omega'}{2\pi} G^A(\omega')\otimes G^R(\omega'+\omega) = 
       \left[-i\omega - i\hat{H}\otimes\mathds{1} + i\mathds{1}\otimes\hat{H}\right.\\
            +\frac{\gamma}{2}\left((1+N)a^\dagger a + Na^\dagger a\right)\otimes\mathds{1}\\ \left. +
              \frac{\gamma}{2}\mathds{1}\otimes\left((1+N)a^\dagger a + Na^\dagger a\right)
            \right]^{-1}.
\end{multline}
The above equation can be proven in time representation by differentiating the product 
$G^A(-t)\otimes G^R(t)$ by $t$.
To get the final expression for $\Pi^<$, one should 
substitute {\eqref{GRGA_integral} into 
\eqref{gamma_1}, \eqref{gamma_2}, 
and then substitute \eqref{gamma_1} and \eqref{gamma_2} into \eqref{polarization_operator}.
Thus, the polarization operator $\Pi^<$ reads
\begin{equation}
    \Pi^< = \Tr\left[\hat{a}^\dagger\{-i\omega \mathds{1} - \mathcal{L}\}(\hat{a}\hat{\rho}_\mathrm{st})\right] + 
            \Tr\left[\hat{a}\{i\omega \mathds{1} - \mathcal{L}\}(\hat{\rho}_\mathrm{st}\hat{a}^\dagger)\right].
\end{equation}
where
\begin{multline}
    \label{lindblad_superoperator}
    \mathcal{L} = -i\hat{H}_0\otimes\mathds{1} + i\mathds{1}\otimes\hat{H}_0^{T}\\
        +\frac{\gamma(1+N)}{2}\left(2\hat{a}\otimes (\hat{a}^\dagger)^T - \hat{a}^\dagger\hat{a} \otimes \mathds{1} - 
        \mathds{1}\otimes (\hat{a}^\dagger \hat{a})^T\right)\\
        +\frac{\gamma N}{2}\left(2\hat{a}^\dagger\otimes \hat{a}^T - \hat{a}\hat{a}^\dagger \otimes \mathds{1} - 
        \mathds{1}\otimes (\hat{a}\hat{a}^\dagger)^T\right).
\end{multline}

The calculations above can be considerably simplified in the limit of small $\gamma$. 
In this limit, the density matrix of the system can be calculated in diagonal approximation
as well as $G^R$ and $G^A$:
\begin{equation}
    \begin{gathered}
        \rho_{nn'} = P_n\delta_{nn'},\\ 
        G^{R,A}_{nn'} = \left(\omega - \epsilon_n \pm \gamma_n\right)^{-1}\delta_{nn'},\\
        \gamma_{n} = \frac{\gamma}{2}[(1+N)\hat{a}^\dagger \hat{a} + N\hat{a}\hat{a}^\dagger]_{nn}.
    \end{gathered}
\end{equation}

When $\omega$ is close to the difference between some energy levels $E_{n} - E_{n'}$, the diagonal approximation
for $G^{R,A}$ can be used for calculating $\Phi_1$, $\Phi_2$, and $\Pi^<$. Retaining only
diagonal retarded and advanced Green functions in Eqs.~\eqref{GRGA_integral}, one can obtain
\begin{equation}
    \int \frac{d\omega'}{2\pi} G^A_{n'n'}(\omega')G^R_{nn}(\omega'+\omega) = 
        \frac{i}{\omega - \epsilon_n + \epsilon_{n'} + i(\gamma_n + \gamma_{n'})}.
\end{equation}
This allows to calculate $\Phi_{1,2}$ in resonant approximation from Eq.~\eqref{phi_1_dyson}, \eqref{phi_2_dyson}:
\begin{equation}
    \begin{gathered}
        (\Phi_1)_{kl} = 
        a^*_{lk}\left(1 + \frac{i\Delta \gamma_{kl}}
            {\omega - \epsilon_k + \epsilon_l - i(\gamma_k + \gamma_l)}\right)^{-1},\\
        (\Phi_2)_{kl} = 
        a_{kl}\left(1 - \frac{i\Delta \gamma_{kl}}
            {\omega - \epsilon_k + \epsilon_l + i(\gamma_k + \gamma_l)}\right)^{-1},\\
        \Delta\gamma_{kl} = \gamma\left((1+N)a_{kk}a^*_{ll} + Na^*_{kk}a_{ll}\right).
    \end{gathered}
\end{equation}

Keeping also only $G^R_{nn}$ and $G^A_{n'n'}$ in $\Pi^<$ given by Eq.~\eqref{polarization_operator}, 
one derives the Eq.~\eqref{lorentz_spectrum} where 
$\Gamma_{nn'} = \gamma_{n} + \gamma_{n'} - \Delta\gamma_{nn'}$.

Thus, the spectrum in the limit of small $\gamma$ consists of multiple narrow Lorentz peaks corresponding to transitions between the system eigenstates.
\end{document}